\documentclass[aps,prc,showpacs,twocolumn,preprintnumbers,floatfix,
showkeys,tightenlines]{revtex4}
\usepackage{amsmath,amssymb,amsfonts}
\usepackage{graphicx}
\usepackage{color}
\allowdisplaybreaks

\newcommand{\be}{\begin{equation}}
\newcommand{\ee}{\end{equation}}
\newcommand{\bea}{\begin{eqnarray}}
\newcommand{\eea}{\end{eqnarray}}

\begin{document}



\title{Determining the density content
 of symmetry energy and neutron skin: an empirical approach}

\author{B. K. \surname{Agrawal}}
\email{bijay.agrawal@saha.ac.in}
\author{J. N. \surname{De}}
\email{jn.de@saha.ac.in}
\author{S. K. \surname{Samaddar}}
\email{santosh.samaddar@saha.ac.in}
\affiliation{
Saha Institute of Nuclear Physics, 1/AF Bidhannagar, Kolkata
{\sl 700064}, India} 


\begin{abstract} 

 The density dependence of nuclear symmetry energy remains poorly
constrained. Starting from precise empirical values of the nuclear
volume  and surface symmetry energy coefficients and the nuclear
saturation density, we show how in the ambit of microscopic
calculations with different energy density functionals, 
the value of the symmetry energy slope parameter $L$ alongwith
that for neutron skin can be put in tighter bounds. The value
of $L$ is found to be $L$= 64$\pm $5 MeV. For $^{208}$Pb, the
neutron skin thickness comes out to be 0.188 $\pm $0.014 fm.
Knowing $L$, the method can be applied to predict neutron skins
of other nuclei.

\end{abstract}



\maketitle

In recent times, there is a cultivated focus on a better understanding
of the density properties of the symmetry energy of nuclear matter.
Particular attention is given to constrain in a narrow window the value
of the symmetry energy slope parameter  
$L$ at the nuclear matter saturation density
$\rho_0$.  In terrestrial context, this parameter  affects the nuclear
binding energies \cite{mol} and the nuclear drip lines and has a crucial
role in determining the neutron density distribution in neutron-rich
nuclei. In astrophysical context, it is also of seminal importance. The
pressure $P_n$ (=$3\rho_0 L $)
 of neutron matter at  $\rho_0$ influences the radii of cold neutron
stars. The cooling of proto-neutron stars through neutrino convection
\cite{rob}, the dynamical evolution of the core-collapse of a massive star
and the associated explosive nucleosynthesis depend sensitively on the
symmetry energy slope parameter \cite{ste,jan}.  
In the droplet model \cite{mye3,mye4}
of the nucleus, the neutron skin is proportional to $L$, a linear
correlation between the neutron-skin thickness of the nucleus and
neutron-star radius \cite{hor} could thus be envisaged.

The symmetry energy slope parameter is defined as
\begin{eqnarray}
 L = 3\rho_0 \frac{\partial C_v(\rho )} {\partial \rho}|_{\rho_0}, 
\end{eqnarray}
where $C_v(\rho )$ is the volume symmetry energy per nucleon of
homogeneous nuclear matter at density $\rho $.  Estimates of $L$
are fraught with much uncertainties. Isospin diffusion predicts $L$
= 88$\pm $ 25 MeV \cite{che,li}, nucleon emission ratios \cite{fam}
favor a value closer to $L \sim $ 55 MeV, isoscaling gives $L \sim
$ 65 MeV \cite{she}. Analysis of giant dipole resonance (GDR) of
$^{208}$Pb \cite{tri} is suggestive of $L \sim $ 45-59 MeV, whereas pygmy
dipole resonance \cite{car} in $^{68}$Ni and $^{132}$Sn would yield an
weighted average in the range $L$= 64.8$\pm $ 15.7 MeV.  Of late, from a
sensitive fit of the experimental nuclear masses to those obtained in the
finite-range droplet model \cite{mol}, the value of $L$ could be fixed in
the bound $L$ = 70 $\pm $15 MeV. Astrophysical observations of neutron
star masses and radii reportedly provide tighter constraints to $L$ to
43 $ < L < $ 52 MeV \cite{ste1}  within 68 $\% $ confidence limits.

Correlation systematics of nuclear isospin with the neutron skin thickness
\cite{cen,war} for a series of nuclei in the framework of the nuclear
droplet model has  been undertaken by the Barcelona Group. This
has yielded a value of $L$ =75 $\pm $25 MeV. The neutron skins were
measured from antiprotonic atom experiments \cite{trz,jas}, systematic
uncertainties involving model assumptions to deal with strong interaction 
is therefore unavoidable. The novel Pb-radius experiment (PREX) at the
Jefferson Laboratory has now been attempted through parity-violation in
electron scattering as a model-independent probe of the neutron density
in $^{208}$Pb \cite{abr}. The neutron skin $R_{skin}= R_n-R_p$ was found to be
0.33 $_{-0.18} ^{+0.16}$ fm, where $R_n$ and $R_p$ are the point neutron
and proton root-mean squared (rms) radii. A reanalysis yielded the value
to be 0.302 $\pm $0.175 fm \cite{hor1}. The droplet model as well as
calculations with class of different interactions, Skyrme or relativistic
mean-field  (RMF), have now clearly established that the neutron skin thickness
of $^{208}$Pb is strongly correlated with the density dependence of
symmetry energy around saturation \cite{bro,typ,fur,tod}. 
In the backdrop of this
information, the large uncertainty in the experimental neutron radius of
$^{208}$Pb seems to be of not much help in putting $L$ in a tighter bound.

Nuclear dipole polarizibility $\alpha_D$ has been suggested \cite{rei,pie}
as an alternative observable constraining the neutron skin.
The recent high resolution $(p,p^\prime )$ measurement \cite{tam}
of $\alpha_D$ yields the neutron skin thickness of $^{208}$Pb to be
0.156$^{+0.025}_{-0.021}$ fm, but the model dependence \cite{pie1}
in the correlation between $R_{skin}$ and $\alpha_D$ assessed in systematic
calculations in the framework of nuclear density functional theory 
is seen to shift the value of $R_{skin}$ to 0.168$\pm $0.022 fm.

 In this communication, we suggest a new method for determining 
the symmetry energy  slope parameter $L$ 
by exploiting the empirical information
on the volume and surface symmetry energy coefficients, $C_v(\rho_0)$
and $C_s$.  The symmetry coefficient
of a finite nucleus $a_{sym}(A)$ can be parametrized as 
\begin{eqnarray}
a_{sym}(A)=C_v(\rho_0)-C_sA^{-1/3}.
\end{eqnarray}
These coefficients  have recently been meticulously studied
\cite{jia} by using the double differences of "experimental"
symmetry energies. This has the advantage that other effects
(such as pairing and shell effects) in symmetry energy can be well canceled
out from the double differences for neighbouring nuclei. The 
correlation between the double differences and the mass number of
nuclei is found to be very compact yielding values of $C_v(\rho_0)$ and
$C_s$ as 32.10 $\pm $0.31 MeV and 58.91 $\pm $1.08 MeV, respectively.
The uncertainties in these symmetry components are much smaller 
than those found earlier.
We show below that these 'experimental' values of $C_v$ and $C_s$ alongwith
empirical information of the proton rms radius 
in a heavy nucleus yields the value of $L$ within narrower limits;
precision information on the neutron skin of 
nuclei  also follows from the analysis.

  We start with the ansatz
\begin{eqnarray}
C_v(\rho )=C_v(\rho_0 )(\frac{\rho }{\rho_0 })^\gamma,
\end{eqnarray}
where $\gamma $
measures the density dependence of the symmetry energy.
In a considerable density range around $\rho_0$  this ansatz is found
to be very consistent with the density dependence 
obtained from the nuclear equation of state (EOS)
with different interactions \cite{che,li,sam} and also from experiments
in intermediate-energy heavy-ion collisions \cite{fam,she}.
At very low densities, however, there are some small deviations.
From Eqs.~(1) and (3),
$L=3\gamma C_v(\rho_0 )$ and the symmetry
incompressibility $K_{sym}=9\rho_0^2 \frac{\partial ^2 C_v(\rho )}
{\partial \rho ^2 }|_{\rho_0} =3L(\gamma -1) $. One can expand the
volume symmetry coefficient around $\rho_0$ as 
\begin{eqnarray}
C_v(\rho) =C_v(\rho_0) \Bigl [1+\sum_{n=1}^\infty \frac{1}{n!} 
\bigl ( \frac{\rho - \rho_0}{\rho_0} \bigr )^n \prod_{k=0}^{n-1}
(\gamma -k) \Bigr ].
\end{eqnarray}
 In the above expansion, keeping terms upto second order has been found
to be a reliable approximation in our subsequent calculations.  
In terms of the symmetry energy slope parameter 
and symmetry incompressibility,
$C_v(\rho )$ is then given as,
\begin{eqnarray}
C_v(\rho )=C_v(\rho_0)-L\epsilon +\frac{K_{sym}}{2} \epsilon ^2 
\end{eqnarray}
where $\epsilon =(\rho_0 -\rho )/(3\rho_0) $. 

 For a finite nucleus of mass number $A$, the symmetry coefficient
$a_{sym}(A)$ is always less than $C_v(\rho_0)$.
The coefficient $a_{sym}(A)$
can be equated to $C_v(\rho_A)$ where $\rho_A$ is an equivalent density, 
always less than $\rho_0$. Using relations (2) to (5), we show
below how $\rho_A$  and 
$\gamma $ (hence $L$ and $K_{sym}$) can be calculated.
From Eqs. (2) and (5)
it follows that 
\begin{eqnarray}
C_sA^{-1/3}=L\epsilon_A~ -\frac{K_{sym}}{2} \epsilon_A^2,
\end{eqnarray}
where $\epsilon_A=(\rho_0-\rho_A )/(3\rho_0)$. From Eq.~(6), 
\begin{eqnarray}
C_s=3C_v(\rho_0)A^{1/3}[\gamma \epsilon_A -\frac{3}{2} 
\gamma (\gamma -1)\epsilon_A^2 ].
\end{eqnarray}
 The value of $C_s$ is an empirically determined
constant (58.91  $\pm $1.08 MeV).

In the local density approximation, 
the symmetry coefficient $a_{sym}(A)$ can be calculated  \cite{sam}
as
\begin{eqnarray}
a_{sym}(A) =\frac{1}{AX_0^2}\int d^3r \rho (r) C_v(\rho (r))
[X(r)]^2,
\end{eqnarray}
where $X_0$ is the isospin asymmetry  (=$(N-Z)/A$) 
of the nucleus, $\rho (r)$ is the sum of the neutron and proton
densities inside the nucleus and $X(r)$ is the local isospin asymmetry. 
The left hand
side  of Eq.~(8) is $C_v(\rho_A)$, hence Eq.~(8) can be 
rewritten as 
\begin{eqnarray}
C_v(\rho_0)(\frac{\rho_A}{\rho_0})^\gamma =\frac{1}{AX_0^2} \int
d^3r \rho (r) C_v(\rho_0)(\frac{\rho(r)}{\rho_0})^\gamma [X(r)]^2.
\end{eqnarray}
Given the neutron-proton density profiles in the nucleus, from Eq.~(9), 
a chosen  value of $\gamma $ gives $\rho_A$ and hence $\epsilon_A$.
The one that satisfies Eq.~(7) is the desired solution for $\gamma $. 
Once $\gamma $
is known, the equivalent density $\rho_A$, the symmetry 
energy slope parameter  and
symmetry incompressibility are determined. 
From Eq.~(7),  there are two solutions for $\gamma$
and hence for $\rho_A$. Calculations show that one of the
solutions is unphysical as this gives $\rho_A \sim 5 \rho_0$.
The procedure so described is expected to work best for heavy
nuclei where volume effects predominate over those coming from surface.
The heavy spherical nucleus $^{208}$Pb usually serves as a benchmark
for extracting nuclear bulk properties, we choose this nucleus
for our calculation.

 Apriori knowledge of $C_v(\rho_0), C_s, \rho_0 $ and the 
proton and neutron density distributions in the nucleus is
required to extract values of $\gamma$ and $\rho_A$.  In different
parametrizations of the nuclear masses \cite{sto,sat,dan1,dan2,mye},
$C_v(\rho_0)$ is $\sim $ 31 MeV, $C_s$ is $\sim $ 55 MeV, and $\rho_0$
hovers around the canonical value $\sim $0.16 fm$^{-3}$ in different
nuclear EOS models. As stated earlier, we
take the values $C_v(\rho_0)$=32.1 $\pm $0.31
MeV and $C_s$ =58.91 $\pm $1.08 MeV. 
This value of $C_v(\rho_0)$ matches very well 
with 32.51 MeV, the one obtained from the latest mass systematics  
by M\"oller $\it et~ al $\cite{mol}; from their quoted value of
28.54 MeV for the surface stiffness parameter $Q$, the value of 
$C_s$ from $^{208}$Pb also comes very close, 58.16 MeV. For saturation
density, we fix $\rho_0$ =0.155 $\pm $ 0.008 fm$^{-3}$; this encompasses
the saturation densities that come out from the EOS of different 
Skyrme and RMF models. The point proton distribution  is
known from experiments, the neutron density distribution is
laced with much uncertainty though. 

 From a  recent covariance analysis \cite{rei},  a lack of correlation
of the neutron skin with some of the fundamental properties of nuclei
like the isoscalar incompressibility, saturation density
and the nucleon effective mass is suggested. 
The binding energy  is also
seen to be a poor isovector indicator. This  is 
in consonance with the suggestion of
effective nucleon-nucleon interactions of different genres,
non relativistic (Skyrme) and relativistic (RMF), that give in
the framework of microscopic mean-field theory different values
of the neutron skin in $^{208}$Pb \cite{bijay,bro} without compromising
the basic nuclear properties mentioned earlier. Aided by  the
further information that the neutron skin 
calculated with an effective interaction is strongly
correlated with the corresponding symmetry 
energy slope parameter $L$ \cite{cen}, with empirical
knowledge of $C_v$, $C_s$, $\rho_0$ and the proton density distribution,
we now show how using Eqs.~(7) and (9) for a heavy nucleus,
both $L$ and $R_{skin}$ can be calculated.

The parameters of the   interactions BSR8-BSR14 \cite{bijay}, 
FSUGOLD  \cite{tod},
NL3 \cite{nl3} and TM1 \cite{sug}  have been used to generate  
the proton and neutron density profiles of $^{208}$Pb
in the RMF model. Alongwith 
many experimentally observed properties of finite nuclei and nuclear
matter, these interactions reproduce the proton r.m.s radius in $^{208}$Pb
($R_p = <r_p^2>^{1/2}$ =5.451 fm) extremely well. The neutron r.m.s
radii  vary considerably though, the  calculated neutron skin varies
from  0.17 fm to  0.28 fm. The symmetry energy slope parameter 
evaluated with these
interactions using Eq.~(1) are  displayed in
Fig.~1 as a function of the corresponding calculated neutron skin $R_{skin}$
(blue filled triangles).  The magnitude of $L$ increases with $R_{skin}$,
its functional dependence $L(R_{skin})$ shows the usual linear correlation.
These different interactions yield, using Eq.~(1),
different values of $C_v (\rho_0 )$ (31-38 MeV).
We also use Eqs.~(7) and (9) to calculate  $\gamma $ and hence $L$
by employing the microscopic densities for the protons
and neutrons obtained  within the RMF models using the above mentioned
parameter sets. These values of $L$ are depicted 
in the same figure by the shaded region,
the spread at a particular $R_{skin}$ arising from the 
uncertainties in the values
of $C_v$, $C_s$ and $\rho_0$. 
\begin{figure}
\resizebox{3.0in}{!}{ \includegraphics[]{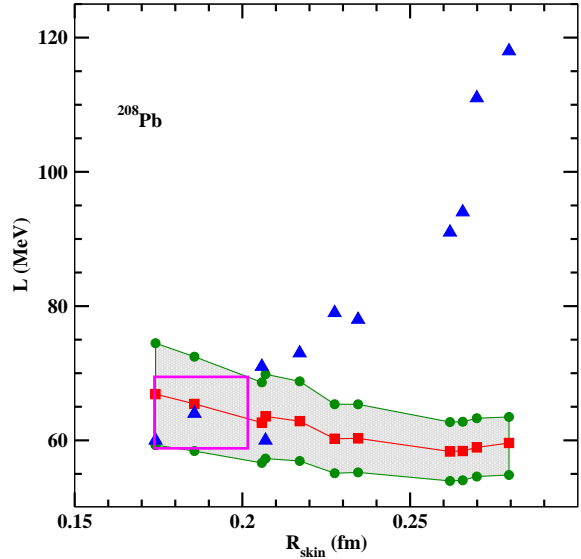}}
\caption{\label{fig:snm} (Color online) The blue triangles represent
 $L$ calculated using Eq.~(1) with different RMF interactions. They are
plotted as a function of corresponding $R_{skin}$ for $^{208}$Pb.
The shaded region represents the envelope of possible $L$-values
with different RMF interactions obtained using Eqs.~(7) and (9).
The acceptable window for the values of $L$ and $R_{skin}$ for 
$^{208}$Pb is represented by the magenta box.
}
  \end{figure}

\begin{figure}
\resizebox{3.0in}{!}{ \includegraphics[]{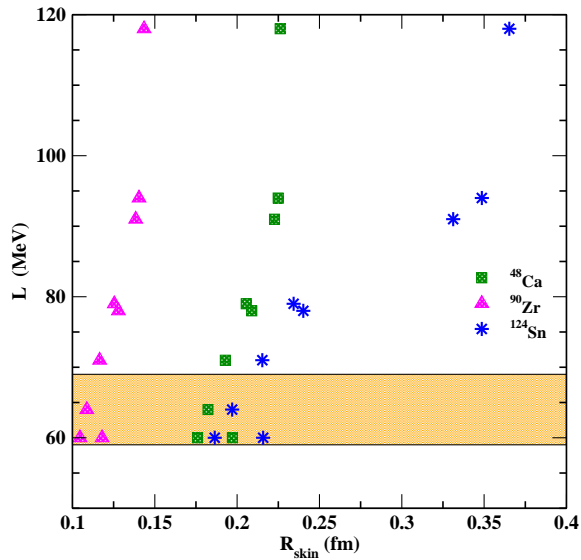}}
\caption{\label{fig:snm} (Color online) The values of $L$ are shown
as a function of $R_{skin}$ for $^{124}$Sn (blue stars), $^{90}$Zr (magenta
triangles) and $^{48}$Ca (green squares) calculated with different 
RMF interactions. The shaded region corresponds to the projected-out range in
the values of $L$ as obtained from Fig.~1.
}
  \end{figure}
The shaded region projects  a 
strikingly different dependence of $L$  on $ R_{skin}$.
The filled red squares in the shade represent the median values for $L$,
the filled green circles represent its lower as well as the upper bounds.
The slope parameter $L$ is seen to decrease
here weakly with the neutron skin. This possibly originates from the
fact that with given values of $\rho_0$, $C_v(\rho_0)$ and $C_s$,
for a particular nucleus, $C_v(\rho_0)(\frac{\rho_A}{\rho_0})^\gamma
=a_{sym}(A)$ (Eq.~(2)) is a fixed quantity, $(\rho_A)^\gamma $ is thus a
constant for all the chosen interactions.  Because $\rho_A$ and $\gamma
$ have to satisfy Eq.~(7), there is not much latitude in their values,
resulting in the  weak variation in $L$ as shown.  The intersection
of the linear function $L(R_{skin})$ with the shaded region projects out those
values of neutron skin for $^{208}$Pb that are commensurate with the
given "experimental" windows for $C_v(\rho_0), C_s$ and $\rho_0$. The
corresponding calculated values of $L$ are also accordingly projected
out. The section depicting the bounds of $L$ and also of the neuron
skin of $^{208}$Pb is shown by the box (magenta)  in the figure. We
find them to be $L =$ 64 $\pm $5 MeV and 
$R_{skin}$($^{208}$Pb) = 0.188 $\pm $ 0.014 fm.  
Our scheme for finding $L$ is found to be quite robust. This
is tested by choosing Woods-Saxon density profiles which are
realistic but may not be very accurate.
The proton density is adjusted to reproduce the experimental
proton rms radius for $^{208}$Pb, 
the neutron density profiles are varied so that
the entire range of 0.13 fm to 0.47 fm for  $R_{skin}$ as obtained from
the PREX experiment is covered. 
Even in this large range of $R_{skin}$, $L$ is confined within
55 to 85 MeV; the median value (70 MeV) is not much different from
the one obtained with microscopic densities. 
Knowledge of $L$ helps in predicting neutron skins of nuclei.
Estimates for the neutron skin for a few nuclei
are displayed in Fig.~2; they
are obtained from the intersection of the shaded region 
showing the calculated limits of $L$ =64 $\pm $5 
obtained from microscopic mean-field densities with the
linear function $L(R_{skin})$ calculated for those nuclei in the RMF model
with different energy density functionals. 
The values of the  neutron skins displayed for the nuclei
$^{124}$Sn, $^{90}$Zr and $^{48}$Ca are seen to be 
0.196$\pm $0.014,
0.107$\pm $0.007 and 0.182$\pm $0.008 fm, respectively.

 In summary, based on the empirical knowledge of the 
volume and surface symmetry coefficients and the nuclear saturation
density, we have presented a model that yields the density dependence
of the symmetry energy in tighter bounds. 
This helps in making
a precision prediction of the neutron skin of  different nuclei,
including the currently experimentally studied nucleus $^{208}$Pb,
in PREX. 
We suggest that our determination on the limits on $L$ and $R_{skin}$ 
of experimentally studied nuclei
be considered in properly evaluating nuclear energy 
density functionals. 

\begin{acknowledgments}
 J.N.D  acknowledges support of DST, Government of India. The 
authors gratefully acknowledge the assistance of Tanuja Agrawal
in the preparation of the manuscript.
\end{acknowledgments}


\begin{thebibliography}{99}

\bibitem{mol} Peter M\"oller, William D. Myers, Heroic Sagawa,
and Satoshi Yoshida, Phys. Rev. Lett. {\bf 108}, 052501 (2012).

\bibitem{rob} L. F. Roberts, G. Shen, V. Cirigliano, J. A. Pons,
S. Reddy, and S. E. Woosley, Phys. Rev. Lett. {\bf 108} 061103 (2012).


\bibitem{ste} A. W. Steiner, M. Prakash, J. M. Lattimer, and P. J. Ellis,
Phys. Rep. {\bf 411},  325 (2005).
 
\bibitem{jan} H.-Th. Janka, K. Langanke, A. Marek, G. Mart\'inez-Pinedo,
and B. M\"uller, Phys. Rep. {\bf 442}, 38 (2007).

\bibitem{mye3} W. D. Myers and W. J. Swiatecki, Ann. Phys. (N. Y.) 
{\bf 55}, 395 (1969).

\bibitem{mye4} W. D. Myers and W. J. Swiatecki, Nucl. Phys. {\bf A336},
267 (1980).

\bibitem{hor} C. J. Horowitz and J. Piekarewicz, Phys. Rev. C {\bf 64},
062802 (2001).

\bibitem{che} L. W. Chen, C. M. Ko, and B. A. Li, Phys. Rev. Lett. {\bf 94},
032701 (2005)

\bibitem{li} B. A. Li, L. W. Chen, and C. M. Ko, Phys. Rep. {\bf 464},
113 (2008).

\bibitem{fam} M. A. Famiano {\it et. al.,} Phys. Rev. Lett. {\bf 97},
052701 (2006).

\bibitem{she} D. V. Shetty, S. J. Yennello, and G. A. Souliotis,
Phys. Rev. C {\bf 76}, 024606 (2007). 

\bibitem{tri} L. Trippa, G. Col\'o, and E. Vigezzi, Phys. Rev. C {\bf 77},
061304 (R) (2008).

\bibitem{car} Andrea Carbone, Gianluca Col\'o, Angela Bracco, Li-Gang Cao,
Pier Francesco Bortignon, Franco Camera, and Oliver Wieland, Phys. Rev. C
{\bf 81}, 041301 (R) (2010).

\bibitem{ste1} A. W. Steiner and S. Gandolfi, Phys. Rev. Lett. {\bf 108},
081102 (2012).


\bibitem{cen} M. Centelles, X. Roca-Maza, X. Vi\~nas, and M. Warda,
Phys. Rev. Lett. {\bf 102}, 122502 (2009).

\bibitem{war} M. Warda, X. Vi\~nas, X. Roca-Maza, and M. Centelles,
Phys. Rev. C {\bf 80}, 024316 (2009).

\bibitem{trz} A. Trzci\'nska {\it et. al.}, Phys. Rev. Lett. {\bf 87},
082501 (2001).

\bibitem{jas} J. Jastrzebski {\it et. al. }, Int. J. Mod. Phys. E {\bf 13},
343 (2004).

\bibitem{abr} A. Abrahamyan {\it et. al.}, Phys. Rev. Lett. {\bf 108},
112502 (2012).

\bibitem{hor1} C. J. Horowitz {\it et. al. }, Phys. Rev. C {\bf 85},
032501(R) (2012).

\bibitem{bro} B. A. Brown, Phys. Rev. Lett. {\bf 85}, 5296 (2000).

\bibitem{typ} S. Typel and B. A. Brown, Phys. Rev. C {\bf 64}, 027302 (2001).

\bibitem{fur} R. J. Furnstahl, Nucl. Phys. A {\bf 706}, 85 (2002).

\bibitem{tod} B. G. Todd-Rutel and J. Piekarewicz, Phys. Rev. Lett. {\bf 95},
122501 (2005).

\bibitem{rei} P.-G. Reinhard and W. Nazarewicz,  Phys. Rev. C {\bf 81},
051303(R) 2010.

\bibitem{pie} J. Piekarewicz, Phys. Rev. C {\bf 83}, 034319 (2011).

\bibitem{tam} A. Tammi {\it et. al.,} Phys. Rev. Lett. {\bf 107},
062502 (2011).

\bibitem{pie1} J. Piekarewicz, B. K. Agrawal, G. Colo, W. Nazarewicz,
N. Paar, P.-G. Reinhard, X. Roca-Maza, and D. Vretenar,
Phys. Rev. C {\bf 85}, 041302 (2012).

\bibitem{jia} H. Jiang, G. J. Fu, Y. M. Zhao, and A. Arima,
Phys. Rev. C {\bf 85}, 024301 (2012).

\bibitem{sam} S. K. Samaddar, J. N. De, X. Vi\~nas, and M. Centelles,
Phys. Rev. C {\bf 76}, 041602(R) (2007).


\bibitem{sto} M. Stoitsov, R. B. Cakirli, R. F. Casten, W. Nazarewicz,
and W. Satula, Phys. Rev. Lett. {\bf 98}, 132502 (2007).

\bibitem{sat} W. Satula, R. A. Wyss, and M. Rafelski, Phys. Rev. C 
{\bf 74}, 011301(R) (2006).

\bibitem{dan1} P.Danielewicz, Nucl. Phys. A{\bf 727}, 233 (2003).

\bibitem{dan2} P. Danielewicz and J. Lee, Nucl. Phys. A{\bf 818}
36 (2009).

\bibitem{mye} W. D. Myers and W. J. Swiatecki, Nucl. Phys. A{\bf 601}
141 (1996).

\bibitem{bijay} B. K. Agrawal, Phys. Rev. C {\bf 76}, 045801 (2007).

\bibitem{nl3} G. A. Lalazissis, J. K\"onig, and P. Ring, 
Phys. Rev. C {\bf 55}, 540 (1997).

\bibitem{sug} Y. Sugahara and H. Toki, Nucl. Phys. {\bf A579},
557 (1994).


\end{thebibliography}
\end{document}